\documentclass{article}

\usepackage{arxiv}

\usepackage[utf8]{inputenc} % allow utf-8 input
\usepackage[T1]{fontenc}    % use 8-bit T1 fonts
\usepackage{hyperref}       % hyperlinks
\usepackage{url}            % simple URL typesetting
\usepackage{booktabs}       % professional-quality tables
\usepackage{amsfonts}       % blackboard math symbols
\usepackage{nicefrac}       % compact symbols for 1/2, etc.
\usepackage{microtype}      % microtypography
\usepackage{lipsum}		% Can be removed after putting your text content
\usepackage{graphicx}
\usepackage{natbib}
\usepackage{doi}
\usepackage{amsmath,amssymb,amsfonts}
\usepackage{algorithmic}
\usepackage{graphicx}
\usepackage{booktabs}
\usepackage{tikz}
\usepackage{pdfrender}
\usetikzlibrary{positioning, arrows.meta}
\usepackage{textcomp}
\usepackage{xcolor}

\title{Anomaly Detection in Cooperative Vehicle Perception Systems under Imperfect Communication}

\author{ 
	Ashish Bastola \\
	School of Computing\\
	Clemson University\\
	Clemson, USA\\
	\texttt{abastol@clemson.edu} \\
	\And
	Hao Wang \\
	School of Computing\\
	Clemson University\\
	Clemson, USA\\
	\texttt{hao9@clemson.edu} \\
        \And
	Abolfazl Razi \\
	School of Computing\\
	Clemson University\\
	Clemson, USA\\
	\texttt{arazi@clemson.edu} \\
}

\renewcommand{\shorttitle}{\textit{Anomaly Detection in Cooperative Vehicle Perception Systems under Imperfect Communication} }

\begin{document}
\maketitle

\begin{abstract}
Anomaly detection is a critical requirement for ensuring safety in autonomous driving. In this work, we leverage Cooperative Perception to share information across nearby vehicles, enabling more accurate identification and consensus of anomalous behaviors in complex traffic scenarios. To account for the real-world challenge of imperfect communication, we propose a cooperative-perception-based anomaly detection framework (CPAD), which is a robust architecture that remains effective under communication interruptions, thereby facilitating reliable performance even in low-bandwidth settings. Since no multi-agent anomaly detection dataset exists for vehicle trajectories, we introduce 15,000 different scenarios with a 90,000 trajectories benchmark dataset generated through rule-based vehicle dynamics analysis. Empirical results demonstrate that our approach outperforms standard anomaly classification methods in F1-score, AUC and showcase strong robustness to agent connection interruptions. The code and dataset will be made publicly available at:https://github.com/abastola0/CPAD
\end{abstract}


\section{Introduction}
\begin{figure}[ht!]
        \centering
        \includegraphics[width=0.8\linewidth]{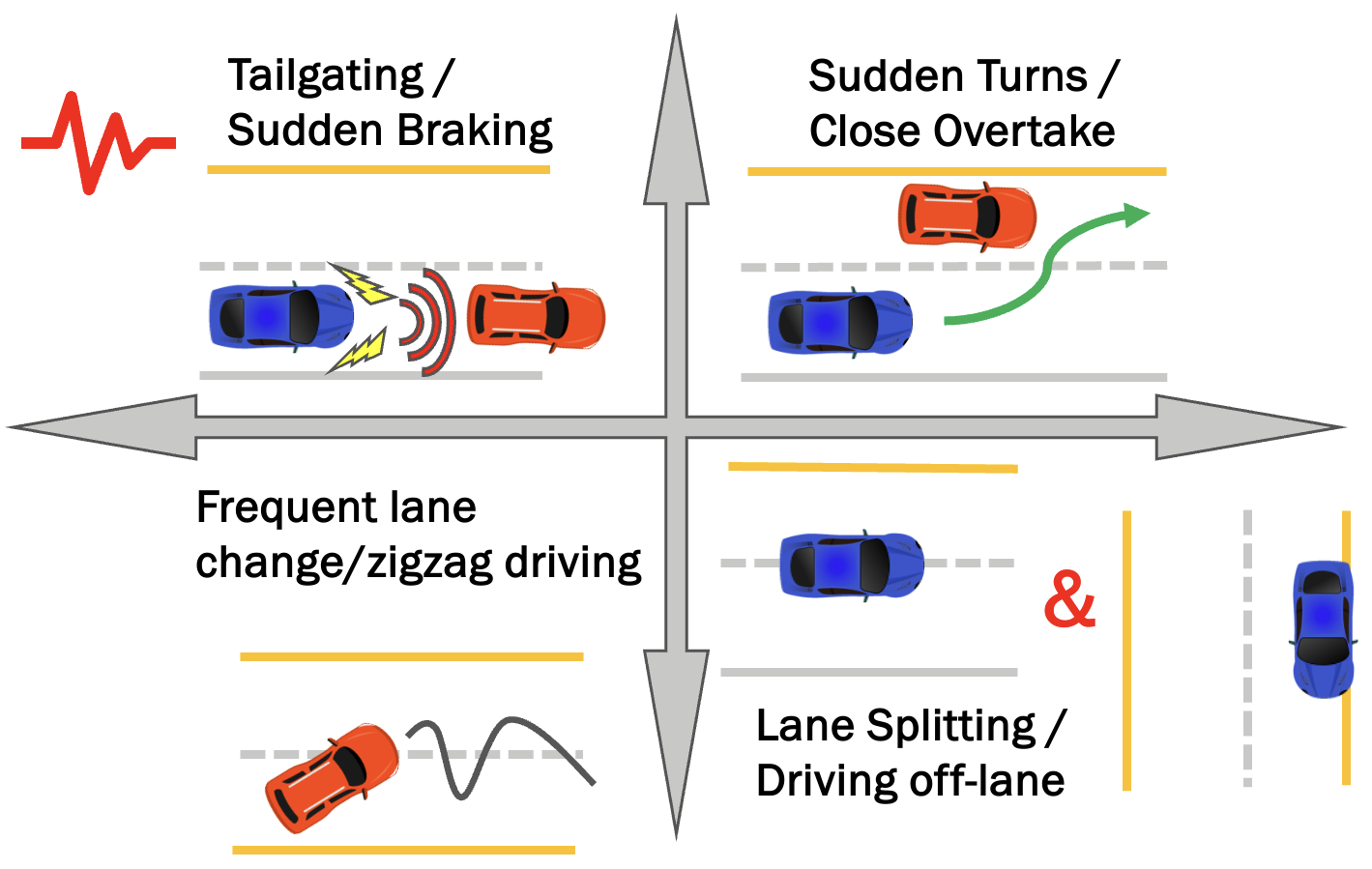}
        \caption{Anomalies classified using rule-based approach}
        \label{fig:anomaly_types}
    \end{figure}

Ensuring safety in autonomous vehicles requires identifying abnormal vehicle behaviors or unexpected traffic patterns ahead of time. To enable large-scale deployment, however, traditional single-vehicle approaches to anomaly detection struggle as they fail to capture the broader context of multi-agent interaction present in real-world driving scenarios. Moreover, despite significant advances in sensor technology in recent years, the perception capability of these local sensors is ultimately bounded in range and field of view (FOV) due to their physical constraints. Cooperative perception(CP) offers a solution by enabling vehicles to share information about their surroundings from their perspectives, thus providing a richer overall view of the environment. Modern vehicle and self-driving equipment manufacturers like Tesla, CommaAI, etc. are increasingly shifting towards pure perception models that rely heavily on front-facing perception systems. The more abstract these sensors become, the risk of failing to detect anomalous behaviors significantly increases. Thus, CP can be beneficial in these scenarios by accessing additional information that a vehicle might miss, with the assistance of other vehicles. One simple instance might be lane changing when another vehicle is on the blindside, as shown in figure \ref{fig:anomaly_types}. Even though many blindside monitors are prevalent these days, these technologies are still immature to some extent, and many might not even have access to these technologies or have limited access depending on the manufacturer. With a CP-based approach, the vehicles observing from behind can help identify this abnormal behavior and report or even override lane change prevention to avoid accidents through a common consensus. Another similar instance might be a vehicle driving between lanes or offset from the center due to sensor malfunction or performing unnecessary and frequent lane changes than required. More complex scenarios might arise where a single vehicle perception is insufficient. For instance, a real-world incident captured a driverless Waymo vehicle driving toward oncoming traffic. This behavior might have been easily prevented if surrounding vehicle perspective information had been incorporated into Waymo’s decision-making system, which could have prevented the vehicle from driving in the wrong direction. Thus, CP provides a point of reference to guide the decision-making process in these kinds of uncertain scenarios.

Despite these benefits, CP-based methods face practical constraints: the requirement for significantly fast real-time inference to avoid delays, appropriate consensus of other agents' information in cases of ambiguity, unreliable communication channels, and bandwidth limitations that can lead to missing information or interrupted connection. We thus propose a robust cooperative-perception-based anomaly detection(CPAD) framework to address these challenges in handling imperfect information sharing. Our method models spatio-temporal correlations among vehicle trajectories to detect behavioral anomalies, even in incomplete data.

\begin{figure*}
    \centering
    \includegraphics[width=1\linewidth]{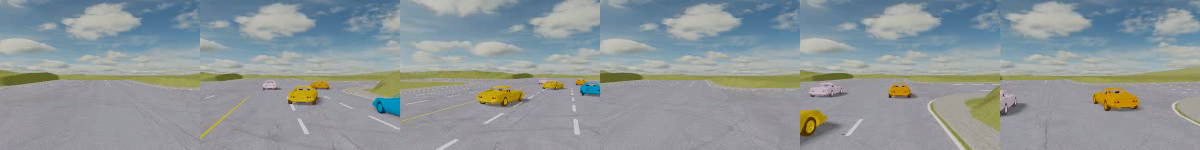}
    \caption{Demonstrates individual perception of vehicles in 6 agent settings. We observe similar perception but in the form of lidars, lane detectors and side detectors instead of visual graphics for computational efficiency.}
    \label{fig:multiagent-visual}
\end{figure*}

Our contribution is threefold:
\begin{itemize}
    \item Since no standard multi-agent anomaly detection dataset for vehicle trajectories currently exists, we introduce a 15,000(6 agent sets with a 90,000 single-agent) multi-agent trajectory benchmark dataset generated through rule-based correlation analysis. 
    \item We propose a robust spatiotemporal cooperative-perception-based anomaly detection (CPAD) architecture based on a graph transformer that can handle missing node communication and temporal agent interactions to detect anomalous behavior. Experimental results demonstrate that our graph-based node modeling framework achieves superior F1 scores and accuracy compared to conventional classification methods. 
    \item Second, we perform a robustness analysis of our model by considering real-world communication interruptions. We perform sequential and stepwise node communication blackouts leading to lost perception and evaluate our model under these circumstances.
\end{itemize}
    Our approach lays the groundwork for safer, more reliable autonomous driving. Our code and dataset will be made publicly available to foster further research in this domain.

\section{Related Works}
    \subsection{Collective information sharing}
        CP has been recognized for its potential to enhance road safety by facilitating the sharing of raw or processed sensor data through vehicular communications\cite{aoki2022time}. Another approach includes sharing latent representation after being processed by a feature extractor\cite{wang2024cmp}. CP systems are particularly beneficial in safety-critical and dense traffic environments, as they can increase the situational awareness and safety of both vulnerable road users (VRUs) and in-vehicle users through integration with roadside units (RSUs) and Vehicle-to-Everything (V2X) network\cite{shan2020demonstrations}. Advances in communication technologies, especially the anticipated capabilities of sixth-generation wireless networks, promise to support the sharing of raw perception data over millimeter wave frequencies\cite{zhang2022joint}. However, the extensive bandwidth required for transmitting raw sensory data and the need for real-time communication still introduce significant challenges. These include bottlenecks in data processing and latency in achieving consensus among nodes. Current approaches to addressing these challenges often involve inference in probabilistic graphical models to facilitate node-wise consensus on predictions\cite{liu2012variational}. Simple approaches like voting system \cite{shan2021discrete, hassanzadeh2013building} are widely adopted, however, their efficacy is still questionable and does not account for temporal dynamics and an agent's confidence. 
    
    \subsection{Graph-based information modeling}
    Graph-based traffic modeling has been widely adopted in traffic forecasting applications. Temporal Graph Convolution neural network (T-GCN) has been used to learn complex topological structures to capture the spatial dependence using Graph Convolution networks (GCN) alongside RNN to capture the temporal dynamics of moving vehicles \cite{zhao2019t}. Graph convolution policy networks based on Reinforcement learning have been used to generate dynamic graph structures when the dynamic graphs are incomplete due to the data sparsity\cite{peng2021dynamic}. GNN-s PePerceptionhave also been used in a self-supervised fashion by leveraging edge features \cite{caville2022anomal}; however, their implementation is limited to just spatial anomaly detection. Approaches for detecting anomalies in dynamic graphs are generally called online anomaly detection approaches\cite{yu2018netwalk, bhatia2020midas, chang2021f}. These approaches accumulate the edges for a certain time period and form a graph snapshot\cite{lamichhane2024anomaly}. However, Anomaly detection using these approaches poses challenges due to growing graph volumes and expanding feature spaces from continuously arriving edges and real-time data reception, respectively. These challenges, along with the single-pass processing requirement, impact time and space complexities and anomaly detection accuracy and thus are infeasible for large trajectory data\cite{lamichhane2024anomaly}. Our method exploits the scaling ability of transformer architecture \cite{vaswani2017attention} to account for long-range temporal correspondence and mean global pooling to extract same-sized vector embedding in dynamic node structure, which allows it to handle large data volumes with significantly low compute for graph abstraction.
    \subsection{Anomaly detection in Communication constrained Scenarios}
    For instance, vehicles in connected systems in regions such as highways are dispersed over a large area with temporally changing proximity. This can induce communication interruptions and disconnect when agents leave their communication range. Using multiple data sources is a long sought-after problem \cite{zhao2006robust}. To account for communication constraints, especially with streaming data, various isolation forests have been proposed with PCA as a preprocessing abstraction \cite{jain2021anomaly}; however fail to account for temporal consistency between sensor data. \cite{alippi2016model}. Generative adversarial neural networks (GAN) based approaches have been used to generate a latent representation of visual data to account for communication constraints\cite{li2020cognitive}; however, they fail to account for joint information sharing between multiple communication nodes. Federated learning approaches are another way for joint training  \cite{mcmahan2017communication, bastola2024fedmil}. Various pruning-based approaches \cite{magnani2023pruning}, and diversity-based client selection approaches \cite{bastola2024fedmil} have been proposed to account for communication bottlenecks. However, these methods generally operate over single client data for each specific sample in a distributed fashion and do not account for joint agent data consensus or temporal correspondence. Thus are completely different than the CP problem we're trying to solve. 
    Based on our knowledge, none of the current works account for all three factors of temporal dynamics, node dynamics and real-time computability at the same time distributed traffic setting. Our approach accounts for this gap and proposes a graph transformer-based architecture that accounts for all of the above factors and scalability to large traffic data and performs significantly well under node blackouts.
\section{Problem Formulation}
\subsection{Multi-agent configuration}
The key problem in this study is the joint classification of anomalous behavior within a group of vehicles that remain under communication range. We first set up a modified version of the metadrive simulator\cite{li2022metadrive} to accommodate multi-agent settings. Each vehicle in this setting has its observation in the form of 240 lasers for lidar and 12 lasers for lane line detectors and side detecters each. These lasers primarily detect moving objects, sidewalks/solid lines etc. In our case, we consider 6 agent settings assumed to be nearby but with occasional or complete disconnect and signal corruption during communication. Our developed model should be able to handle these variable agent scenarios of missing agents or even the addition of new agents, as well as be robust against corrupted signals.
The next problem we solve is the trajectory labeling. Defining anomalous trajectories is not straightforward as it involves accounting for factors such as sudden lane change, tailgating, sudden braking etc. As per our knowledge, there are no anomaly detection datasets or benchmarks to facilitate more focused research in this area. We thus design a mixture of rule-based trajectory labelers with human-validated thresholds to categorize each trajectory as anomalous or not. This dataset contains 90,000 different multi-agent trajectories for each agent in 15,000 scenarios that can be easily incorporated to train for specific anomaly types. These trajectories are divided into 75\% normal and 25\% anomalous labels. We also provide the code to record and relabel these trajectories based on any desired rule-based anomaly labeling.

    \subsection{Rule-based anomaly labeling}
    We assume we're only given the perception information in the form of Lidar, lane and side detectors. We make this assumption to accommodate for the fact that these are fundamental sensors available to almost any autonomous vehicle one can imagine, and countless other sensors might serve additional functionalities to these sensors. However, since we can access vehicle dynamics information in the simulator, we exploit it to generate targets for trajectory anomalies. During inference we rely on just the perception information to reason if any trajectory is anomalous. Following are the individual anomalies that we detect in our labeler:
    \subsubsection{Zigzag driving patterns}

    Given the vehicle heading $\theta_x$ and $\theta_y$, we detect the curvature of trajectory at any given time as follows:
    \begin{equation}
    %\small
        k=\frac{\left|\theta_x^{\prime} \theta_y^{\prime \prime}-\theta_y^{\prime} \theta_x^{\prime \prime}\right|}{\left(\theta_x^{\prime 2}+\theta_y^{\prime 2}\right)^{3 / 2}}
    \end{equation}

    where $\theta_x^{\prime}=\frac{d \theta_x}{d t}, \theta_y^{\prime}=\frac{d \theta_y}{d t}, \theta_x^{\prime \prime}=\frac{d^2 \theta_x}{d t^2}, \theta_y^{\prime \prime}=\frac{d^2 \theta_y}{d t^2}$. We compute these time derivatives based on the temporal window size of 40 and threshold this value based on human observation to detect high-frequency zig-zag patterns.
    \subsubsection{Sudden Braking}
    We calculate the velocity magnitude $V_i=\left\|\mathbf{v}_i\right\|$ for every time point $i$. Acceleration can thus be calculated using finite differences over a specified window size given by: 
    $A_i=\frac{S_{i+w / 2}-S_{i-w / 2}}{w}$ where $w$ is the window size and $A_i$ is the acceleration at point $i$. We then compute the smoothed acceleration using a moving average filter as $S A_i=\frac{1}{w} \sum_{k=-\frac{w-1}{2}}^{\frac{\mathrm{m}-1}{2}} A_{i+k}$. Here, the index $k$ runs between the half-windows $-\frac{w-1}{2}$ to $\frac{w-1}{2}$, which ensures that the window is centered at $i$ and includes $w$ terms. This formula assumes that $w$ is odd; if $w$ is even, you might need to adjust the limits accordingly or redefine the window slightly to maintain symmetry.

    Then we identify sudden acceleration changes when $\text{Indices}=\left\{i \mid S A_i<\theta\right\}$, where $\theta$ represents the braking threshold which is typically a negative value indicating deceleration.
    
    \subsubsection{Sudden turns}
    Similarly, measuring lateral acceleration is the best way to detect sudden turns. Given a sequence of heading vectors $\mathbf{h}_i$ (where $i$ indexes the frame), the angular change $\Delta \theta_i$ between consecutive heading vectors can be calculated as: $\Delta \theta_i=\arccos \left(\frac{\mathbf{h}_{i-1} \cdot \mathbf{h}_i}{\left\|\mathbf{h}_{i-1}\right\|\left\|\mathbf{h}_i\right\|}\right)$, where $\mathbf{h}_{i-1}$ and $\mathbf{h}_i$ are normalized to unit vectors before computing the dot product. We clip values to ensure the argument of arccos remains within its valid domain of $[-1,1]$ to prevent undefined values due to floating-point precision issues. The lateral acceleration $a_{\text {lat }, i}$ based on the angular change and velocities is computed using:
    $$
    %\small
    a_{\mathrm{lat}, i}=\Delta \theta_i \times v_{i+1}
    $$
    where $v_{i+1}$ is the magnitude of the velocity vector at frame $i+1$, and $\Delta \theta_i$ is the angular change from the previous frame. To identify frames where the lateral force exceeds a certain threshold $\tau$ (such as a critical lateral acceleration that might indicate a risk of skidding or rolling in vehicle dynamics), we use:
    $$
    \text { Indices }=\left\{i| | a_{\text {lat }, i} \mid>\tau\right\}
    $$

Here, $\tau$ is set to $0.8 \mathrm{~m} / \mathrm{s}^2$ but could be adjusted based on empirical data or dynamic conditions.
    \subsubsection{Frequent lane switching/driving between two lanes}
    We detect this specific lane anomaly using lane break points. We trigger a boolean whenever a vehicle crosses a lane line and evaluate the crossing frequency to detect anomalous vehicle trajectory. The moving average gives a better estimate for this as well. The moving average, $M$, of the data array $D$, is computed using a convolution operation. This operation involves sliding a window of size $w$ across $D$, and for each position, computing the average of the elements within the window given by:
    $M=D * \frac{1}{w} \mathbf{1}_w$, where, $\mathbf{1} _w$ is a vector of ones with length $w$, $*$ denotes the convolution operation and $\frac1w\mathbf{1}_w$ is the kernel used for the convolution, effectively averaging the elements in each window of size $w$. This is central to smoothing the data, reducing noise, and helping to identify sustained patterns indicative of lane changes.
    We then identify indices where $M$ exceeds 0.5 :
    $$
    \text { Indices }=\left\{i \mid M_i>0.5\right\}
    $$

    We derive contiguous index ranges as Intervals and then filter these by computing the length of each interval based on a threshold $\tau$. 
    \begin{equation}
    %\small
        \mathrm{Anomalies}=\{\ell(\mathrm{Int.})\mid\mathrm{Int.}\in\mathrm{Intervals~}\wedge\ell(\mathrm{Int.})>\tau\}
    \end{equation}
Here, $\ell(\mathrm{Int.})$ is the length of a specific interval and $\tau$ is the user-validated threshold indicating significant duration indicative of an anomaly in lane change behavior.

    \subsubsection{Tailgating/ Driving very close}
    We use the simulator-provided proximity detectors to detect if any vehicles exist close to a specific threshold to detect these behaviors.
    
\begin{figure*}
    \centering
    \includegraphics[width=1\linewidth]{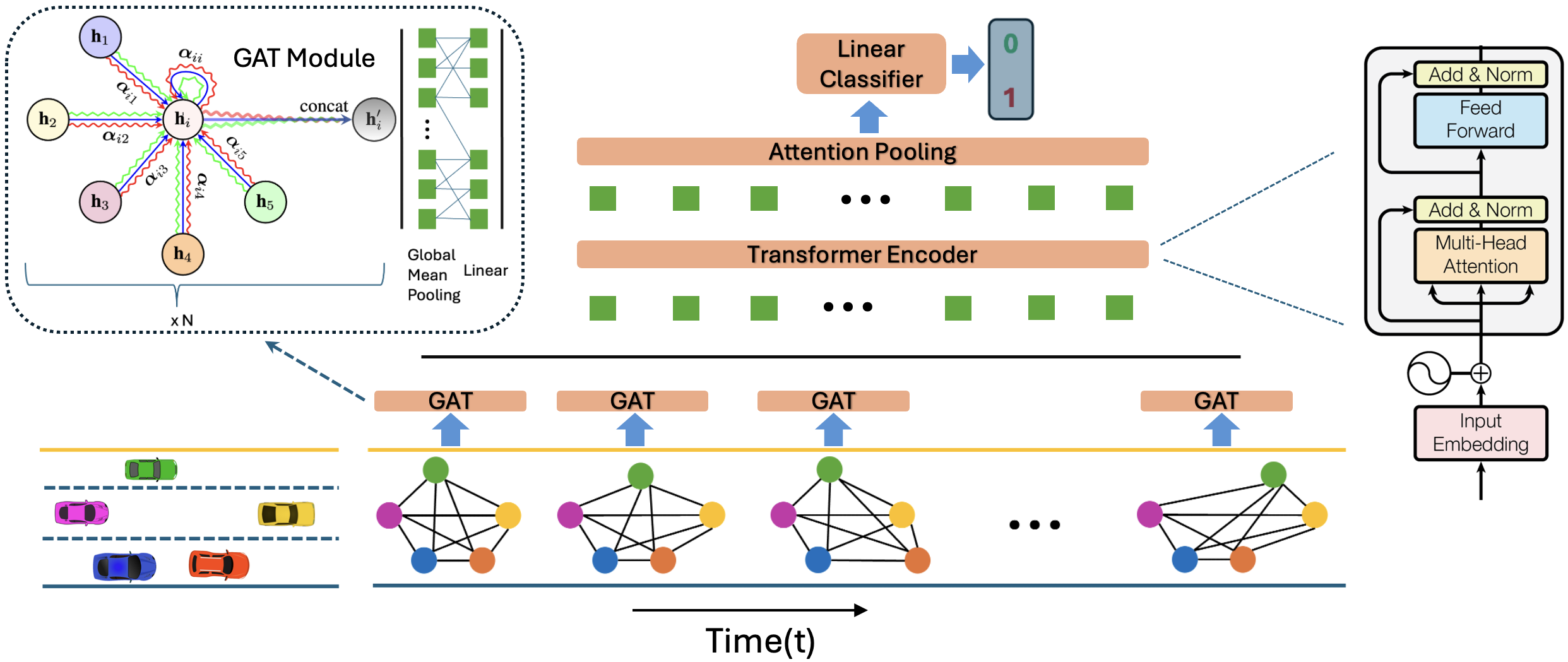}
    \caption{Graph-Transformer Architecture of the Cooperative Perception-Based Anomaly Detection (CPAD) Model for Trajectory Anomaly Detection.}
    \label{fig:architecture}
\end{figure*}
\begin{figure}
    \centering
    \includegraphics[width=1\linewidth]{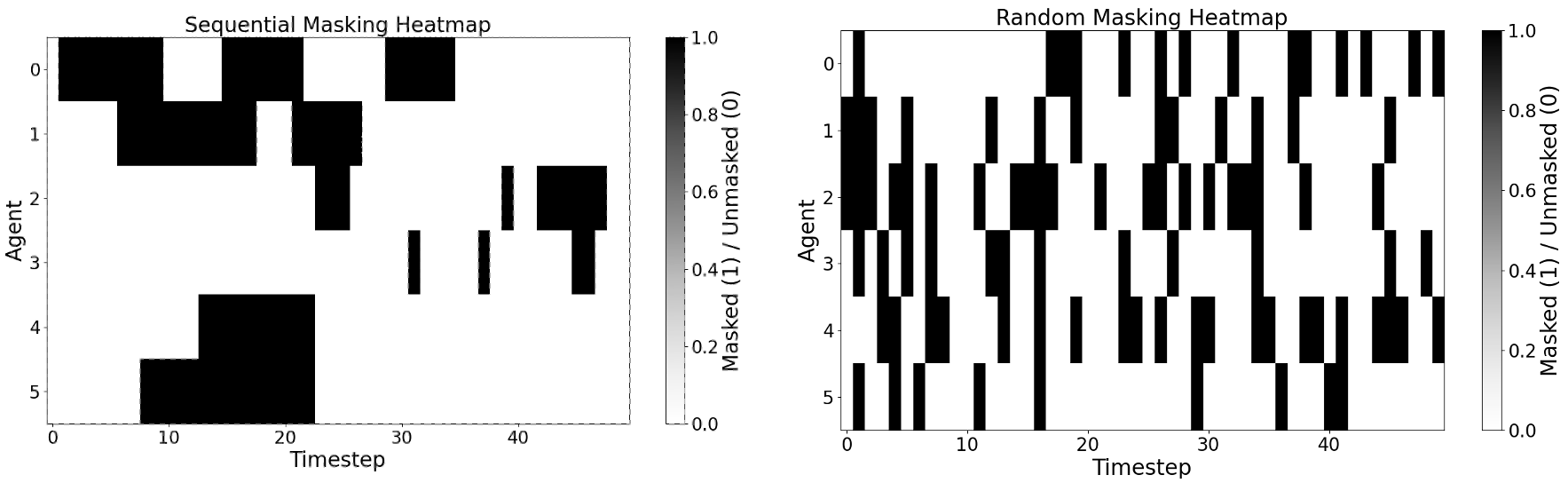}
    \caption{Node blackout configurations. Sequential blackout on the left where the node goes out of communication for a certain period of time. Random blackout on the right where the node goes through intermittent connection loss.}
    \label{fig:agent-masking}
\end{figure}
\section{Algorithm Design}
\subsection{Graph based Spatial sensor fusion} \label{sub:graph}
Multi-agent systems, in general, focus on the interactions and decision-making processes among autonomous agents to maximize common objectives, which may or may not involve CP. In our case of CP, however, we aim to passively detect anomalous behavior using the perception information from multiple vehicles, trying to maximize the correct predictions. One note to this integration is the dynamic number of agents. Even though the Deep learning approach for this variation in nodes can be handled with several padding-based architectures, graph-based methods are the widely used and most effective ones\cite{velickovic2017graph, wang2019heterogeneous}. We thus first employ graph-based abstraction to capture the spatial correlation between dynamic nodes. Let $\mathbf{h}_i^{(m)}$ be the sensor readings vector of vehicle $i$ with dimension $m$. The node features linear transformation is given by:
$$\mathbf{h}_i^{\prime}=\textbf{W}^{(m \times n)}\mathbf{h}_i^{(m)}$$
where $\mathbf{W}$ is a weight matrix of hidden dimension $n$, applied to every node, and $\mathbf{h}_i^{\prime}$ is the transformed feature.
We then compute the attention coefficients $e_{ij}$ that indicate the importance of node $j$'s features to node $i$ as:
\begin{equation}
    %\small
    \mathbf{e}_{ij}=\mathrm{LeakyReLU}\left(\mathbf{a}^T[\mathbf{h}_i^{\prime}\|\mathbf{h}_j^{\prime}]\right)
\end{equation}

Here, a is a weight vector, and $\|$ denotes concatenation. These attention coefficients are then normalized using softmax across all choices of $j$ in the neighborhood $\mathcal{N}_i$ of node $i$, making $\alpha_{ij}$ effectively the weight of the influence of node $j$ on node $i$.
\begin{equation}
    %\small
    \boldsymbol{\mathbf{\alpha}}_{ij}=\mathrm{softmax}_j(\mathbf{e}_{ij})=\frac{\exp(\mathbf{e}_{ij})}{\sum_{k\in\mathcal{N}_i}\exp(\mathbf{e}_{ik})}
\end{equation}
Multi-head attention is then applied, and their features can now be concatenated (denoted by $\|$) to form the final output features for each node. Our method worked well with multi-head attention concat compared to average as we stack two GAT layers. When we do so, the output of each GAT will be heads($K$) $\times$ the hidden dimension($n$). The final embedding can thus be computed as follows:
\begin{equation}
    %\small
    \mathbf{h}_i^{(K\times n)}=\left\|_{k=1}^K\sigma\left(\sum_{j\in\mathcal{N}_i}\alpha_{ij}^{(k)}\mathbf{h}_j^{\prime(k)}\right.\right)
\end{equation}
We use multiple of this Graph transformation followed by ReLU activation for each stack. The final layer after stacking is the linear layer followed by the global mean pooling layer given by:
\begin{equation}
%\small
    \mathbf{h}_G = \mathbf{W^\prime} \left(\frac{1}{N} \sum_{i=1}^N \mathbf{hp}_i\right) + \mathbf{b}, where; 
    \mathbf{hp}_i = \frac{1}{N} \sum_{i=1}^N \mathbf{h}_i
\end{equation}

This computation is also demonstrated in figure \ref{fig:architecture}. A major benefit of this graph-based abstraction module is its ability to handle a changing number of nodes and fully connected information sharing between nodes. This helps derive information from other agents if it's unavailable to that specific agent.

\begin{table*}[htpb]
\centering
\caption{Comparision of CPAD with various Anomaly Detection methods}
\label{tab:my-table}
\resizebox{0.8\textwidth}{!}{%
\begin{tabular}{@{}lllllll}
\toprule
\textbf{Models} & \textbf{F1-Score} & \textbf{AUC} & \textbf{Precision} & \textbf{Recall} & \textbf{MCC} & \textbf{Accuracy} \\ \midrule
LOF               & 0.01& 0.62& 1.0& 0.0& 0.06& 0.80\\
LSTM& 0.55& 0.82& 0.49& 0.62& 0.42& 0.79\\
Decision Tree               & 0.20& 0.56& 0.77& 0.11& 0.25& 0.81\\
SP-AE               & 0.53& 0.85& 0.71& 0.42& 0.47& 0.85\\
Ours**            & \textbf{0.70}& \textbf{0.90}& 0.78& \textbf{0.63}& \textbf{0.64}& \textbf{0.89}\\ \bottomrule
\end{tabular}%
}
\end{table*}

\begin{figure*}[ht!]
    \centering
    \includegraphics[width=1\linewidth]{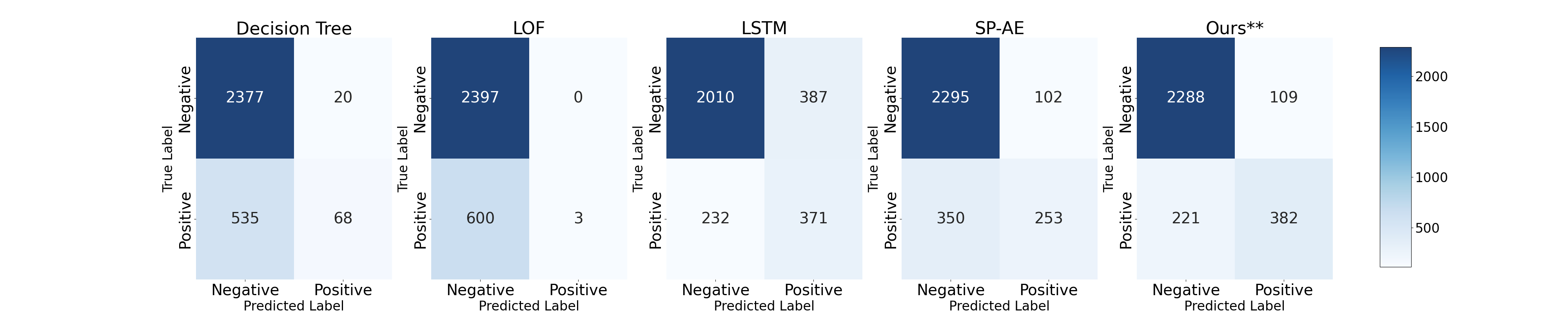}
    \caption{Confusion Matrix of True vs. Predicted Anomaly Labels}
    \label{fig:Confusion Matrix}
\end{figure*}

\begin{figure}
    \centering
    \includegraphics[width=0.7\linewidth]{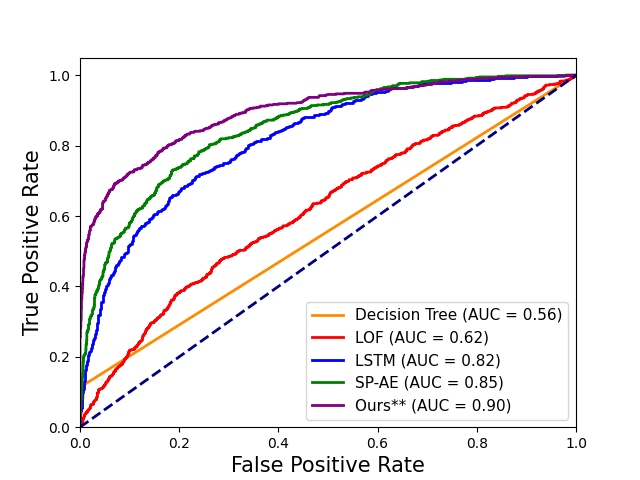}
    \caption{ROC curve demonstrating the performance comparison for un-masked trajectory inference.}
    \label{fig:roc_curve}
\end{figure}
\subsection{Sequential Binary Classifier}
For each timestep $t$, we do the Graph-based spatial sensor fusion as in \ref{sub:graph} to generate a final sequence $\mathbf{h}_G^0, \mathbf{h}_G^1, \ldots, \mathbf{h}_G^T$. This sequence is then fed to the transformer encoder along with position encoding to generate a final latent representation $\mathbf{h}_G'^{0:T}$ as follows:
\begin{equation}
    %\small
    {\mathbf{h}_G'^{0:T}} = \text{Transformer}(\mathbf{h}_G^0, \mathbf{h}_G^1, \ldots, \mathbf{h}_G^T) 
\end{equation}
The sequence $\mathbf{h}_G'^{0:T}$ is passed through an attention pooling layer, which computes a weighted sum of the features, where the weights are learned attention scores given by :
    \[
    %\small
    \text{AttentionPooling}(\mathbf{x^\prime}) = \mathbf{z}= \sum_{t=0}^T \text{softmax}(\mathbf{w_{ap}}^\top \mathbf{x^\prime}_t) \cdot \mathbf{x^\prime}_t
    \]
    where $\mathbf{x^\prime}_t$ represents each vector in the sequence $\mathbf{h}_G'^{0:T}$, and $\mathbf{w_{ap}}$ is the learned weight vector from a linear transformation applied to each $\mathbf{x^\prime}_t$. The pooled output is then fed into a binary classifier, which outputs the final prediction:
    \[
    %\small
    \text{Classifier}(\mathbf{z}) = \sigma(\mathbf{W}_c \mathbf{z} + b_c)
    \]
    where $\mathbf{z}$ is the prediction logit from the attention pooling, $\sigma$ denotes the sigmoid activation function, $\mathbf{W}_c$ and $b_c$ are the weights and biases for the classifier. Sigmoid is applied to this logit, and a threshold of 0.5 is used to make the binary class prediction to decide anomaly.

\begin{table*}[htpb]
\centering
\caption{Robustness to Random stepwise blackouts}
\label{tab:step-blackout}
\resizebox{0.8\textwidth}{!}{%
\begin{tabular}{@{}lllllll}
\toprule
\textbf{Blackout Percentage} & \textbf{F1-Score} & \textbf{AUC} & \textbf{Precision} & \textbf{Recall} & \textbf{MCC} & \textbf{Accuracy} \\ \midrule
2\%& 0.702& 0.895& 0.828& 0.609& 0.652& 0.896\\
5\%& 0.691& 0.893& 0.832& 0.590& 0.642& 0.894\\
8\%& 0.690& 0.891& 0.846& 0.582& 0.645& 0.895\\
 10\%& 0.682& 0.886& 0.844& 0.572& 0.637&0.893\\
15\%& 0.661& 0.881& 0.864& 0.536& 0.624& 0.890\\
25\%& 0.605& 0.857& 0.875& 0.463& 0.580& 0.879\\ \bottomrule
\end{tabular}%
}
\end{table*}
\section{Experiments}
\subsection{Baseline Comparisions}
We show that our proposed Graph-Transformer architecture performs better than some of the most popular anomaly classifiers like LOF\cite{breunig2000lof}, LSTM\cite{hochreiter1997long}, Decision Tree\cite{quinlan1986induction} and supervised-autoencoders(SP-AE)\cite{le2018supervised}. We evaluate our model in terms of F1-Score, AUC, Precision, Recall and Matthews correlation coefficient(MCC), given the imbalanced nature of our training data. Figure \ref{fig:roc_curve} shows the ROC performance comparison.

\subsection{Robustness under imperfect communication}
The second set of experiments includes random step-wise and sequential node blackouts. 
These simulate real-world communication interruptions or packet loss. The results demonstrate that our model can perform significantly well up to 10\% of communication loss with just a negligible loss on the F1-score while still maintaining 0.68 under step-wise node blackouts. Our model performs well even under 25\% information loss, still maintaining an F1-score above 0.6 as shown in the table \ref{tab:step-blackout}. 

The sequential blackout occurs in random timestep blocks with a maximum block size of 10 timesteps. This case is more extreme than random blackouts as it overlaps most anomalous trajectory segments and contributes to major information loss. Our model performs significantly well even in this scenario for up to 15\% intensity while maintaining an F1-score above 0.6. These results are shown in table \ref{tab:seq-blackouts}

These experiment results indicate a significant robustness of graph-based training for cooperative information sharing among nodes. 
%A multi-agent vehicle setting becomes even more intriguing when there is imperfect communication between nodes. 
%Our model can achieve an F1 score of 0.7 under the threshold of 0.5. This score is higher as the dataset we use for training is highly unbalanced and without undersampling or oversampling. 

The SP-AE is derived from the training strategy similar to \cite{le2018supervised} where we combine reconstruction loss and classification loss as an auxiliary guide.

%\begin{table}[htpb]
%\centering
%\caption{}
%\label{tab:my-table}
%\resizebox{0.5\textwidth}{!}{%
%\begin{tabular}{@{}lllllll}
%\toprule
%\textbf{Agent Perception} & \textbf{F1-Score} & \textbf{AUC} & \textbf{Precision} & \textbf{Recall} & \textbf{MCC} & \textbf{Accuracy} \\ \midrule
%A1               & 0.01& 0.62& 1.0& 0.0& 0.06& 0.80\\
%A2& 0.55& 0.82& 0.49& 0.62& 0.42& 0.79\\
%A3               & 0.20& 0.56& 0.77& 0.11& 0.25& 0.81\\
%A4               & 0.53& 0.85& 0.71& 0.42& 0.47& 0.85\\
%A5               & 0.53& 0.85& 0.71& 0.42& 0.47& 0.85\\
%A6               & 0.53& 0.85& 0.71& 0.42& 0.47& 0.85\\
%Majority Voting               & 0.53& 0.85& 0.71& 0.42& 0.47& 0.85\\
%Averaging               & 0.53& 0.85& 0.71& 0.42& 0.47& 0.85\\
%Weighted Voting               & 0.53& 0.85& 0.71& 0.42& 0.47& 0.85\\
%Ours**            & \textbf{0.70}& \textbf{0.90}& 0.78& \textbf{0.63}& \textbf{0.64}& \textbf{0.89}\\ \bottomrule
%\end{tabular}%
%}
%\end{table}
%

\begin{table}[htpb]
\centering
\caption{Robustness to Sequential blackouts}
\label{tab:seq-blackouts}
\resizebox{0.8\textwidth}{!}{%
\begin{tabular}{@{}lllllll}
\toprule
\textbf{Blackout Percentage} & \textbf{F1-Score} & \textbf{AUC} & \textbf{Precision} & \textbf{Recall} & \textbf{MCC} & \textbf{Accuracy} \\ \midrule
2\%& 0.699& 0.894& 0.833& 0.602& 0.652& 0.896\\
5\%& 0.688& 0.891& 0.831& 0.587& 0.640& 0.893\\
8\%& 0.652& 0.879& 0.829& 0.537& 0.606& 0.885\\
 10\%& 0.652& 0.875& 0.843& 0.532& 0.611&0.886\\
15\%& 0.615& 0.865& 0.858& 0.479& 0.583& 0.879\\
25\%& 0.575& 0.845& 0.869& 0.430& 0.554& 0.872\\ \bottomrule
\end{tabular}%
}
\end{table}

\section{Conclusion}
In our study, we introduced a robust multi-agent framework for anomaly detection in cooperative vehicle perception systems under imperfect communication conditions. Our proposed Graph-Transformer architecture, complemented by a newly curated benchmark dataset of 15,000 vehicle trajectories, demonstrates superior performance in identifying and classifying anomalous behaviors in complex traffic scenarios, surpassing conventional anomaly detection methods. We perform the empirical evaluations that underscore the resilience of our model against various communication disruption types, maintaining high detection accuracy even with up to 25\% communication losses. The open availability of our dataset and codebase aims to foster further advancements and facilitate real-world applications in autonomous driving technologies specifically for anomaly detection. Future work will explore scaling this framework to accommodate larger fleets and integrate more diverse sensor modalities, enhancing the robustness and applicability of CP systems.

\bibliographystyle{plainnat}
\bibliography{references} 

\end{document}